# Mobility Management as a Service for 5G Networks

Akshay Jain, Elena Lopez-Aguilera and Ilker Demirkol
Dept. of Network Engineering
Universitat Politècnica de Catalunya
Barcelona, ES 08034
Email: akshay.jain@upc.edu, elopez@entel.upc.edu, ilker.demirkol@entel.upc.edu

*Abstract*—Mobility Management (MM) techniques have conventionally been centralized in nature, wherein a single network entity has been responsible for handling the mobility related tasks of the mobile nodes attached to the network. However, an exponential growth in network traffic and the number of users has ushered in the concept of providing Mobility Management as a Service (MMaaS) to the wireless nodes attached to the 5G networks. Allowing for on-demand mobility management solutions will not only provide the network with the flexibility that it needs to accommodate the many different use cases that are to be served by future networks, but it will also provide the network with the scalability that is needed alongside the flexibility to serve future networks. And hence, in this paper, a detailed study of MMaaS has been provided, highlighting its benefits and challenges for 5G networks. Additionally, the very important property of granularity of service which is deeply intertwined with the scalability and flexibility requirements of the future wireless networks, and a consequence of MMaaS, has also been discussed in detail.

## I. INTRODUCTION

The provision of mobility is what makes the current day wireless networks an indispensable quantity in our daily lives. Without mobility management, every time users would have to either forfeit their services or buy a new SIM card when moving to another geographic location. However, through mobility management continuity of services is ensured, which also enhances the Quality of Experience (QoE) for the users attached to the network.

Current mobility management architectures, such as the one employed by LTE [1], are centralized in nature. To illustrate, the Mobility Management Entity (MME) in the LTE architecture depicted in Fig. 1 is the central entity which is entrusted with the responsibility of managing mobility of users attached to the network. The aforementioned central architecture suffices current day needs. However, due to an exponential growth in traffic and the number of users [2], these architectures will not be viable for the future 5G network scenarios. Issues such as scalability and flexibility, which are intricately connected to the granularity of service aspect, will render the current strategies insufficient for the scenarios that will prevail in these future networks.

As a consequence of the realization of above mentioned issues, Software Defined Networking (SDN) and Network Function Virtualization (NFV) have been recognized as important enablers for the future networks. Through the softwarization of networks, not only do SDN and NFV help in reducing the Capital Expenditure (CAPEX)/Operating Expenditure (OPEX) for the network operators, but they also enable the implementation of critical network functions, such as mobility management, as applications on top of a central/distributed controller. And so, with a global/locally-global perspective of the complete network architecture, the mobility management applications employed will be able to provide an on-demand service to the user devices. It must be mentioned here that the aforementioned global perspective relates to the scenario when the employed MM application has the complete network view, whilst a locally-global perspective indicates that the employed MM application has a global perspective of only a specific domain, which, for example, may be a geographical area that the SDN-controller, on which it is employed, covers. Further, the granularity of service provided by mobility management applications will also equip the networks with pre-requisites such as flexibility and scalability necessary to meet the demands of the 5G networks. Henceforth, in this article we discuss in detail the features of this on-demand mobility management service, better known as MMaaS, as well as the related granularity aspects along with their benefits and challenges for 5G networks.

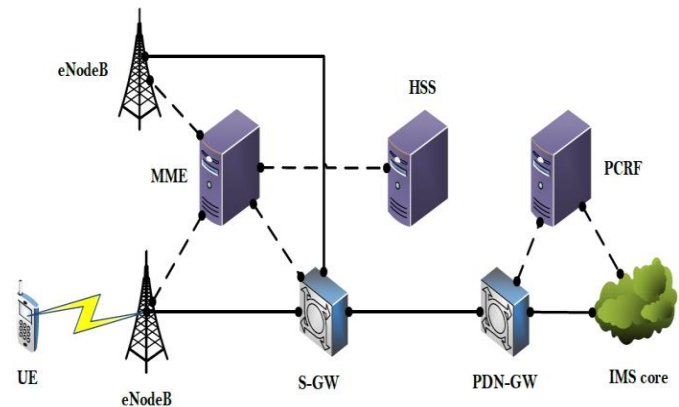

Fig. 1. LTE Architecture.

However, before discussing the concept of MMaaS in detail, it is important to note that surveys such as [3] and [4] provide an important basis for understanding the challenges and opportunities that will exist when implementing current/legacy mobility management solutions in a dense and heterogeneous network environment, such as the 5G wireless networks. While they provide detailed analysis on the many efforts that have been made to support seamless mobility, this article extends the aforementioned studies by providing significant insights into the new paradigm of MMaaS. Further, in [5] the SDN and NFV techniques have been implemented not just at the network side, but also at the mobile node (MN) and the correspondent node (CN). A policy based mobility management technique on a per-flow basis has been proposed. While, through the proposed technique a flexible and granular mobility management strategy (on the basis of flows) has been proposed, our work studies multiple other avenues (such as mobility profiles, flow types, network load and predefined policies) where granularity of service can be offered under the MMaaS paradigm. Additionally, certain projects such as 5G-NORMA [6] provide for an SDN and NFV enabled, and self adjusting mobility framework. Further, [6] also proposes granularity of service based on the required cell size (which indirectly connects to the mobility of the user and the required quality of service), as well as discusses the on-demand mobility management for different slices and sessions. In contrast, not only does this article build upon the aforementioned approaches to mobility management, but it also provides for a detailed description on the various advantages, challenges and the distinct avenues for a flexible mobility management strategy for 5G networks under the MMaaS paradigm.

And hence, with this background, to the best of our knowledge this work is unique in providing a detailed study with regards to MMaaS, and the multiple avenues for granularity in the provision of mobility management services. The remainder of this paper is structured as follows: Section II discusses the concept of MMaaS in more detail. It also provides for a comparison between MMaaS and the current/legacy approaches to mobility management. Section III then describes the benefits of granularity of service, and also presents the various avenues for providing on-demand mobility management. Section IV expresses the challenges that will be faced in the design and implementation of MMaaS. This paper is then concluded in Section V where a broad summary of MMaaS, granularity of service and their benefits and challenges, have also been provided.

II. MOBILITY MANAGEMENT-AS-A-SERVICE (MMAAS)

As mentioned in Section I, and from Fig. 1, current and legacy mobility management strategies have primarily been centralized in nature. But, with the SDN and NFV techniques, mobility management functionalities can now be implemented as an application on top of a controller that provides it with a global view of the domain it serves. This softwarized control over mobility management permits the operators to provide the services on-demand, i.e., MMaaS. It is worth noting that, under the current mobility management strategies when an MN attaches to the network a mobility instance is created for it in the MME, and is kept at all times until it de-registers from the network. This leads to the unnecessary utilization of computational resources. However, with MMaaS, mobility management instances can be created on-demand and hence, computational resources can be allocated likewise. Consequently, MMaaS, through its global view and on-demand computational resource allocation, enables the provision of globally optimized solutions for managing user mobility.

The aforementioned softwarized control allows for the utilization of a versatile set of parameters, which not only provide a globally optimal solution but also permit the self-adjustment of the established mobility management mechanisms. In order to retrieve these parameter values from the network entities, a network controller, i.e., the SDN-controller (SDN-C), has to interact with these entities over the southbound interface (SBI) and then pass on the extracted values to the mobility management application over the northbound interface (NBI) [7]. An illustration of the aforementioned process is provided through Fig. 2. As can be seen from Fig. 2(a), the SDN-C is connected to the OpenFlow (OF) switches which comprise the network data plane. These switches are additionally also connected to the access network, from where values of the parameters such as Signal to Noise Ratio (SNR)/Received Signal Strength Indicator (RSSI) of other and current access points (APs) at the MN, types of flows on the MN, MN policies, etc., can be enquired. Further, from the OF switches, information related to the network such as network load, link failure/congestion information, as well as the latency over the links, etc., can be extracted. All of this information, is then processed at the SDN-C which then, as is visible in Fig. 2(a), is sent over the NBI to the mobility management application. These mobility management applications, which may be implemented on a software cloud, after processing this data, provide a solution to the SDN-C which implements it over the network via the orchestrator through the SBI. Fig. 2(b) provides a signaling diagram to illustrate the above flow of information to ensure mobility management services to the MNs attached to the network.

It is important to mention here that the message sequence as provided in Fig. 2(b) might change in practical implementation. However, the overall logical flow, i.e., information enquiry $\rightarrow$ information reception $\rightarrow$ information processing $\rightarrow$ MM rule implementation, is maintained. Further, in Fig. 2(a), the access network consists of Centralized/Cloud RAN (C-RAN) [8], which is primarily composed of a Baseband Unit (BBU) pool and multiple APs attached to this BBU pool. In the aforementioned scenario, the BBU pool is responsible for handling the access network mobility, i.e., handling the mobility of MNs when they switch APs within the same BBU domain. Here, a BBU domain specifically refers to a set of APs controlled by a particular BBU pool. And so as a consequence, the resource allocation rules message, as shown in Fig. 2(b), is sent by the SDN-C to the access network only when the scenario demands, for instance: when performing a traffic transfer. Thus, it can also be inferred that MMaaS is essentially distributed wherein the access network mobility is handled at the BBU pool whilst network level mobility (inter-domain mobility) is handled at the SDN-C. Lastly, an important point of consideration with regards to the resource allocation rules procedure is that, in the event legacy RAN deployments exist, the SDN-C, similar to the MME as shown in Fig. 1, will almost always be in communication with the access network to handle the mobility at the access network level.

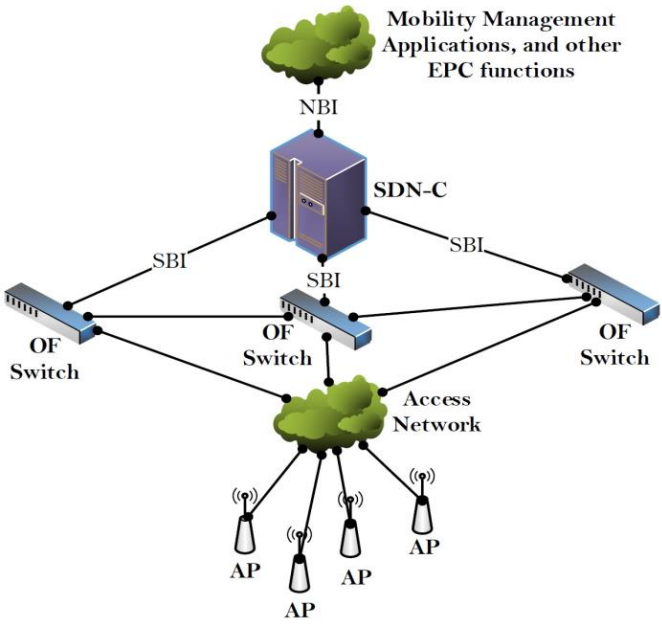

(a)

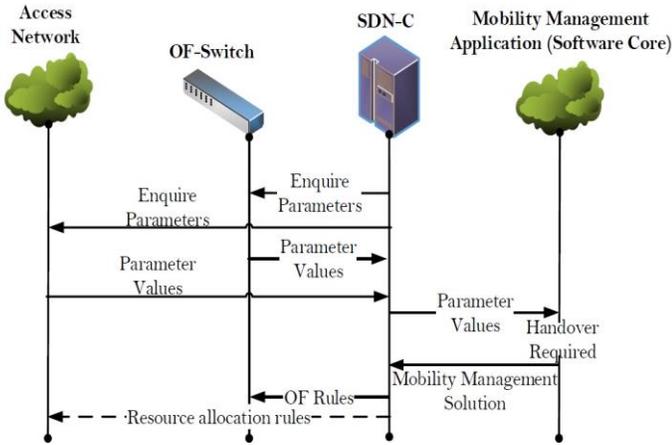

(b)

Fig. 2. (a) Softwarized network control; (b) Signaling diagram for mobility management in SDN based networks.

Next, in order to further exemplify the advantages that MMaaS provides, a short comparative analysis with respect to the existent/legacy mechanisms has been provided in Table I. From Table I, it is evident that the current mobility management mechanisms, designed and developed by IEEE, IETF and 3GPP, do not provide a significant level of service granularity. Whilst these mechanisms provide at best per-MN granularity in service, MMaaS on the other hand has the ability to provide multiple avenues of granularity such as those based on mobility profiles, flows, policies, MN, etc. It is imperative to state here that, in order to provide the level of flexibility and scalability to the future networks, as would be needed to serve the complex scenarios that will be prevalent, such multi-avenue provision in granularity for mobility management services is an indispensable feature. In addition to the multiple avenues of granularity in service, MMaaS provides a significant advantage over the current/legacy mechanisms by allowing for a de-centralized implementation of mobility management applications. Such flexibility stems from the softwarized control which is a consequence of the SDN and NFV framework. Also, MMaaS through its softwarized control and global view will be able to serve multiple network slices whilst current mechanisms, according to Table I, cannot support network slicing environments. This is so because, current/legacy networks are primarily equipped to handle users accessing voice and broadband services, with Narrow-Band (NB) IoT having only recently been standardized. Lastly, the self-organizing capabilities, i.e., the ability to re-structure the routing rules and the access network resource allocation (if needed) depending on the context of operation, are of prime importance to the future networks as the highly dynamic environment will require the mobility management mechanisms to adapt their solutions according to the scenario without any perceivable latency. MMaaS, owing to its flexibility and granularity characteristics as already mentioned, offers a high degree of self-organizing capabilities. On the contrary, existing solutions as shown in Table I, offer minimal self-organizing features. And hence, this also reinforces the belief that existing MM mechanisms are not well-suited to handle the challenging scenarios that future networks will have to encounter.

TABLE I. COMPARISON BETWEEN MMAAS AND CURRENT/LEGACY ARCHITECTURE

|  | MMaaS | 3GPP | IEEE | IETF |
|---|---|---|---|---|
| **Granularity of service** | Multiple avenues[1] | Per-MN | Per-MN | Per-MN[2] |
| **Degree of Centralization** | De-centralized | Mostly Centralized[3] | Centralized | Centralized[4] |
| **Network Slicing Support** | Yes | Minimal[5] | No | No |
| **Self-reorganizing capabilities** | Very High | Minimal[6] | Minimal | Minimal |

From the analysis so far, it is evident that granularity of service offers significant benefits to the network, through its provision of scalability and flexibility, as well as to the users, through the provision of optimal mobility management solutions dependent on their context. Further, there are multiple avenues where granularity in mobility management services can be offered under the MMaaS concept. And so, in the subsequent section, a detailed study on granularity of service and the various avenues, such as mobility profiles, flow types, network load and policies, where the granularity can be offered has been provided.

---

[1] Per-flow, per-mobility profile, policy based, per-MN, etc.
[2] Multi-path TCP (MPTCP) and Stream Control Transmission Protocol (SCTP) allow for multiple paths/flows. However, IETF does not provide per-flow mobility management in these protocols as of yet.
[3] X2 handovers in LTE offer some form of de-centralization.
[4] IETF DMM working group presents certain studies on distributed frameworks. However, there are no standards level RFC as of now.
[5] Recently NB-IoT has been standardized, which utilizes LTE bands to serve IoT devices.
[6] LTE X2 interface allows some level of self-organizing capabilities

## III. GRANULARITY OF SERVICE

As stated in the previous section, granularity of service is beneficial to the network through its provision of scalability and flexibility, whilst for the users it formulates optimal mobility management solutions, which in turn benefit them by helping reduce the power consumption as well as improve their perceived Quality of Service (QoS). To better elaborate these aforementioned broad benefits, we consider the scenario as specified in Fig. 3. The scenario specified in Fig. 3, is a typical mobility scenario wherein an MN migrates from one AP to another, and also switches its access router (AR) in the process. Further, and referring to Fig. 3 again, at AR-1 the MN has a certain set of active flows. After moving from AR-1 to AR-2, the MN keeps its current flows active and also initiates other services which consequently result in the creation of new flows. It is imperative to note here that AR-1 and AR-2 are merely data-plane (DP) entities and henceforth, in the architecture mentioned in Fig. 2(a), they are equivalent to the OF-switches. Next, whilst switching APs and ARs, the mobility management application analyses parameters and policies which are relevant to both the user and the network. It also checks the context, i.e., the mobility profile, the flow types, etc., and makes a handover/traffic transfer decision.

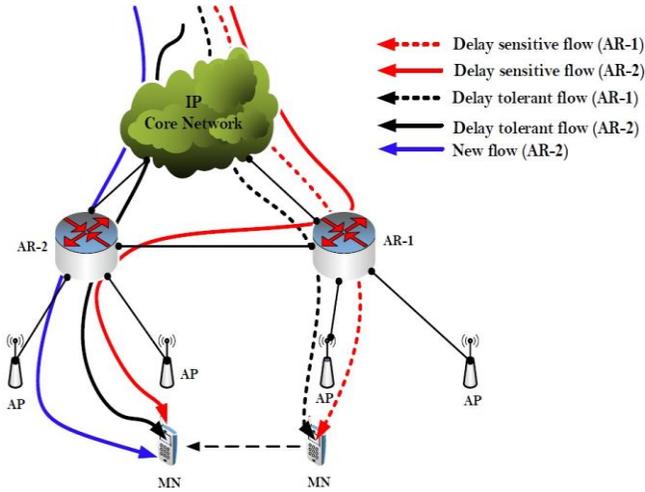

Fig. 3. MMaaS – Granularity of Service provision example.

From the aforementioned decision process, it is clear that the mobility management rules implemented for the MN can provide granularity in terms of mobility profiles, flows, policy, and network load, or a combination of them depending on the context. As an example, in Fig. 3, the granularity of service from the perspective of flows is illustrated, wherein delay-sensitive and delay tolerant flows are served individually with different MM rules. A more detailed discussion for the same is provided in Section III.B.

Such a holistic and distributed decision process provides the network with the flexibility and also allows it to scale itself as the services offered are on-demand and not centralized to just one entity within the complete network for each user. Further, the decision making process involves analyzing the parameters as well as the context. This enables the network to make an optimized decision on mobility management for each user. From the discussions so far, a deeper inspection into the granularity considerations reveals that there are multiple avenues to provide such discretization in mobility management services. Consequently, in the ensuing discussions granularity of service from the perspective of mobility profiles, flow types, network load and predefined policies (both network and user) has been explored in more detail.

### A. Mobility profile perspective

Users in a network can have varying mobility profiles. While most users will be pedestrians, moving at speeds below 3 km/h, some users such as those in a car or high speed trains might be moving at anywhere between 30-500 km/h. Further, with the future networks slated to support Internet of Things (IoT), support for devices with negligible mobility alongside the aforesaid mobility profiles becomes an important point of consideration for future mobility management solutions. The MMaaS paradigm allows operators to deploy softwarized solutions on top of the core network controller, hence, permitting the network to employ solutions that are tailored for specific mobility profiles. As for example, there may be a scenario where there are several pedestrian users with a few high speed users. These users are then overlaid with a high density of static sensors. In such a scenario, the current networks would assign a mobility profile to every attached device, even if they do not require one (like the static sensors). Further, current networks will also employ the same computational and physical resources for each device attached to the network for the purpose of mobility management. While such a method is simple and easy to implement, it is inefficient as certain device do not require similar levels of mobility management, such as static sensors as already mentioned, compared to the others.

Henceforth, MMaaS provides the opportunity to offer granularity on the basis of mobility profiles, wherein devices based on their mobility profile are allocated appropriate resources within the network. Considering the example as described above, a viable solution based on the MMaaS paradigm would be to assign macro-cell resources to the high speed users in order to avoid unnecessary handovers. Further, pedestrians can be allocated control-plane (CP) resources at the macro-cell (to avoid frequent messaging with the core network for resource allocation upon handovers) whilst the data plane can be kept at the small-cells. It is similar to the phantom cell strategy proposed in [9] wherein MNs have CP at the macro-cell and are subjected to a data shower from the small cells, i.e., DP is at the small cells. Additionally, static sensors, since they are not going to be subjected to any mobility event, do not necessitate a mobility profile. And hence, MMaaS avoids assignment of a mobility profile and subsequently, allocation of network resources to such static users. And so, MMaaS ultimately presents a very resource-efficient and flexible avenue, dependent on the user mobility profile, to employ mobility management services.

### B. Flow perspective

With the smartphone boom, and their myriad capabilities, users have access to a variety of applications ranging from the erstwhile calling and short message services (SMS) to the more recent Voice-over-IP (VoIP) services. The unprecedented growth in Internet traffic, which is expected to achieve a rate of 10-12 times by 2021 [2], in conjunction with the ever increasing diversity of the application types has warranted a re-think on how mobility management mechanisms will be able to deal with such heterogeneity.

From Table I, we know that the existent and legacy techniques do not provide for a per-flow granularity. However, with MMaaS, where the mobility management application has a global view of its domain, MM techniques have the capability to distinguish whether a particular application flow is delay sensitive or delay tolerant. Subsequently, upon the determination of the type of flows associated with a user, mechanisms such as allowing data forwarding and the eventual route optimization process for delay sensitive services; and simple IP switching for delay-tolerant services, can be executed by the mobility management applications. To illustrate the aforementioned capability, consider Fig. 3 wherein the MN at AR-1 has two flows. A deeper inspection reveals to the network that one of the flows is delay tolerant while the other is delay sensitive. And so upon moving to a new AP under a new router AR-2, the MMaaS paradigm allows the network to provide data forwarding capabilities for the delay sensitive flow whilst simple IP switching is provided for the delay tolerant flow. Additionally, through MMaaS, the new flow originating at AR-2 is provided access to the IP core network through AR-2 and not AR-1, hence, removing any DP anchoring similar to legacy methods.

Further, applications such as Extreme mobile broadband (xMBB), Ultra-reliable machine-type communications (uMTC), and Massive machine-type communications (mMTC) [10], can be classified as delay-sensitive and delay-tolerant based on their latency requirements, which also encompasses their critical nature. And so, MMaaS through its ability to process each flow separately, as described above, can serve these aforesaid application types satisfactorily ensuring the required network flexibility and the expected QoS for the user, as defined under the 5G paradigm.

*C. Network load perspective*

Network load perspective in essence involves transfer of traffic which implicitly invokes the mobility management mechanisms. Network intelligence, which is a critical component of future networks, allows the transfer of traffic to some other location thus enabling the network to prevent congestion whilst still ensuring the required QoS to the user. Since, this transfer of traffic involves switching the connectivity of user, through MMaaS, the network provides appropriate mobility management rules dependent on the context. As for example, consider the scenario where users in a particular area are in the coverage of multiple APs, i.e, the APs have overlapping coverage areas, and the users can be connected to multiple APs at any given point in time, i.e., they can experience multi-connectivity. Next, these APs might belong to the same RAT or to different RATs, thus leading to a heterogeneous and dense environment. In such a scenario, and given that the density of users is exponentially increasing as mentioned before, some APs or ARs might experience heavy load thus degrading the QoS of the users attached to those network entities.

Henceforth, in order to ease the load on the aforementioned network entities, the network initiates mobility management mechanisms. Subsequently, through the flow and mobility profile based approaches, MMaaS equips the network with the required granularity to transfer certain flows to other points of attachment (in the multi-connectivity scenario) or to completely switch/forward traffic flows depending on their nature. And so, the MMaaS paradigm provides benefits not only to the users, but also extends multiple utilities to the network as discussed above.

*D. Predefined policies perspective*

The mobility management mechanisms implement a particular solution by not just analyzing the parameters and context, but they take into consideration the predefined policies of the network and the user as well. The predefined policies may entail features such as network preference, service subscription, roaming policies, etc. Subsequently, depending on the context of the user, MMaaS can decide to give more weight to certain aspects or more formally: *certain components* of the policy vector, as compared to others. This property essentially enables the mobility management mechanisms to provide specific services to individual users depending on their context with respect to the defined policy vectors. In this regard, [5] and [11] propose ideas for mobility management mechanisms based on policy vectors and user context.

To elaborate, [5] proposes an SDN approach on both the multi-mode mobile terminal (MMT) as well as in the core network. The MMT with assistance from the network gathers information such as available APs and their APIDs, RSSI, network load, etc. This enables the MMT to compare these registered parameter values against its pre-defined policy vectors. Subsequently, it enables the MMT to perform network selection which is then communicated to the core network. The core network then through its SDN-C, and in co-ordination with the C-RAN, orchestrates the required operations in order to provide resources to the MMT over its selected set of APs as well as within the core network. On the other hand, [11] firstly allows the MN to implement its policies when determining the APs it can attach to. After informing the core network about its choice of APs, the network implements its policies and then prunes the list of APs further. The core network subsequently informs the MN about the AP it should attach to. And hence, through the MMaaS paradigm, policy based granularity of service can also be extended towards the users thus emphasizing the utility of MMaaS.

IV. CHALLENGES

Despite the immense benefits of MMaaS and the associated provision of granularity in services, there are multiple challenges that remain to be tackled before MMaaS can be adopted as an enabler for the networks in the near and distant future. In this paper a short discussion on some of these challenges has been provided as follows:

- **Computational Resource Management:** The amount of computational resources that have to be allocated for specific users dependent on their context and profile is critical to the future wireless networks to meet the goals as stated in the 5G paradigm. This is so, because the density of the network and its dynamism will entail that there will be a proper mix of users with different application, mobility and policy profiles. And hence, managing the amount of computational resources to be allocated to each user, i.e., computational resource management, in such an environment will be a challenge.

- **Computational complexity:** MMaaS involves multiple levels of data analysis and optimization of several parameters in order to achieve fairness, guaranteed QoS to user/flow, as well as ensuring network flexibility and scalability. Whilst, many suggested algorithms achieve optimum solutions, time complexity of such methods is always an area of deep concern in real-time scenarios. As already mentioned, since the network will have a high degree of dynamism, ensuring manageable time complexity of the implemented algorithms is a big challenge.

- **Control Plane latency:** While MMaaS does provide the opportunity to control the network entities by implementing OF rules as well as the ability to query them for parameter values, an important point of consideration and ostensibly a challenge will be the latency in the CP. The latency in CP will determine how well the network can serve the dynamic scenarios that will be prevalent in the future networks.

- **Network slicing support:** As already discussed in this article before, service to multiple slices is an important pillar of the 5G paradigm that the future mobility management algorithms will also have to support. MMaaS, will need to be able to support different service types, which will further include joint mobility management solutions or solutions which cater to each individual slice separately [6]. Such diversity in the provision of MM services will be challenging to develop and implement given the diversity in the types of parameters that need to be considered for an optimum mobility management solution for each user.

## V. CONCLUSION

MMaaS through its software control, global view, and on-demand service will be an important enabler for the future wireless networks, i.e., the 5G networks. Through its granularity of service provisions, as studied in detail in this paper, it provides the networks with the flexibility and scalability features so as to cater the highly dense, heterogeneous and dynamic environments that the 5G networks will encounter. Additionally, in this paper, with the help of certain scenarios, the multiple avenues where granularity of services can be provided have been explored in detail, and subsequently, their advantages have been presented as well. Lastly, an analysis of the challenges that the design, development and implementation of MMaaS will encounter has also been provided in this paper. The discussion on the challenges also provides insights on the opportunities that exist for future work on mobility management for 5G networks.

And so, to conclude, MMaaS although faced by multiple challenges will become an important pillar for the future wireless networks, thus enabling them to provide features such as low latency, high data rates, multi-slicing, etc., which are importantly also a part of the broader 5G objectives.

ACKNOWLEDGMENT

This work has been supported in part by the EU Horizon 2020 research and innovation programme under grant agreement No. 675806 (5GAuRA), and by the ERDF and the Spanish Government through project TEC2016-79988-P, AEI/FEDER, UE.